\documentclass[12pt]{article}

\usepackage{epsfig}
\usepackage{amssymb}
\title{Dispersal of Galactic Magnetic Fields into Intracluster Space}

\author{Yinon Arieli, Yoel Rephaeli \& Michael L. Norman}

\def\aa{{\it Astron. Astrophys.} \,}
\def\apj{{\it Ap. J.} \,}

\def\apjl{{\it Ap. J. Lett.} \,}
\def\mn{{\it MNRAS} \,}
\def\aj{{\it Astron. J.} \,}

\def\ea{\ et al. \,}

\def\araa{{\it Annu. Rev. Astron. Astrophys.} \,}
\def\aa{{\it Astron. Astrophys.} \,}
\def\apj{{\it ApJ.} \,}

\def\apjl{{\it Ap. J. Lett.} \,}
\def\mn{{\it MNRAS} \,}
\def\aj{{\it Astron. J.} \,}

\newcommand{\beq}{\begin{eqnarray}}
\newcommand{\eeq}{\end{eqnarray}}

\begin{document}
\title{Dispersal of Galactic Magnetic Fields into Intracluster Space}

\newcommand{\D}{\displaystyle}
\pagenumbering{arabic}
\bibliographystyle{unsrt}
\maketitle


* Accepted for Publication in The Astrophysical Journal

\begin{abstract}

Little is known about the origin and basic properties of magnetic fields 
in clusters of galaxies. High conductivity in magnetized interstellar 
plasma suggests that galactic magnetic fields are (at least partly) 
ejected into intracluster (IC) space by the same processes that 
enrich IC gas with metals. We explore the dispersal of galactic fields 
by hydrodynamical simulations with our new {\em Enzo-Galcon} code, 
which is capable of tracking a large number galaxies during cluster assembly, 
and modeling the processes that disperse their interstellar media. Doing so 
we are able to describe the evolution of the mean strength of the field and 
its profile across the cluster. With the known density profile of dispersed 
gas and an estimated range of coherence scales, we predict the spatial 
distribution of Faraday rotation measure and find it to be consistent with 
observational data. 

\end{abstract}


\section{Introduction}

Direct evidence for magnetic fields in intracluster (IC) gas comes 
from measurements of extended radio emission in many (tens) of clusters, 
and from Faraday rotation measurements in several clusters (reviewed by 
Ferrari \ea 2008, Carilli \& Taylor 2002). Estimates  of the mean strength 
of the field across the radio emitting region are not very reliable because 
of the need to assume a relation between the field and particle (both 
electrons and protons) energy densities. The validity of the commonly made 
assumption of energy equipartition in extragalactic environment of clusters 
is questionable. When equipartition is assumed, field values in the range 
O(0.1-1 $\mu$G) are typically deduced. A reliable lower limit on the mean 
strength of the field, typically $0.2-0.4$ $\mu$G, is obtained when an 
upper limit is set on nonthermal X-ray emission from the radio-emitting 
electrons (reviewed by Rephaeli \ea 2008).

Generally, IC magnetic fields could either arise from cosmological seed 
fields generated in the early universe (reviewed, e.g., by Grasso \& 
Rubinstein 2001) that are directly amplified by turbulent processes 
(e.g., Kunz \ea 2010), or from dispersal of fields from the cluster 
galaxies (Rephaeli 1988). Irrespective of whether very weak primordial 
fields can indeed be efficiently amplified to the above range, the 
possibility of a major contribution to IC fields by magnetized galactic 
plasma should be further explored. The motivation to do so is quite 
clear: IC gas has a major galactic component, as clearly indicated by 
the fact that the gas is metal-enriched. When metals were transferred 
out of galaxies - by galactic winds and ram-pressure stripping - 
interstellar fields embedded in the plasma were carried along. In this 
scenario essentially all the cluster ellipticals (i.e., most of the 
galaxies in a rich cluster) have contributed to the build up of IC fields. 
A significant contribution could have come from a few active galaxies, 
such as dominant radio galaxies (e.g., the Coma cluster; Wilson 1970), 
or AGN (as has recently been simulated by Xu \ea 2009), but due to the 
very uncertain statistics of active galaxies in clusters this contribution 
is difficult to quantify. On the other hand, the contribution of cluster 
ellipticals can be readily gauged by their related metal feedback into IC 
gas. 

Aside from general expectation and simple estimates, a detailed 
description of the stripping of magnetized interstellar (IS) media and 
their direct implications on IC magnetic fields has not yet been given. 
The process is obviously too involved to be described analytically, but 
can be simulated numerically with a hydrodynamical code that has the 
capability of tracking the cluster galaxies. We have recently enhanced 
the powerful adaptive mesh refinement (AMR) hydrodynamical {\em Enzo} code 
(Bryan \& Norman 1997) by adding an algorithm to identify and track a 
relatively large number of galaxies during the course of cluster evolution. 
In our expanded code a galaxy is identified early in the proto-cluster 
evolution, and described subsequently in terms of a galaxy construct, or 
{\it galcon}. Results from the first implementation of our {\em Enzo-Galcon} 
code to model the ejection and transport of gas and metallicity into IC space 
by ram-pressure stripping and galactic winds were described by Arieli, Rephaeli 
\& Norman (2008, 2010, hereafter ARN1, and ARN2). To assess the viability 
of possibly dominant galactic origin of IC magnetic fields, we used this 
(non-MHD) {\em Enzo-Galcon} to describe the ejection of magnetized plasma 
out of {\it galcons}, and determined the spatial distribution and evolution of 
IC fields. In this paper we describe the simulations and the deduced 
properties of IC fields. 

In Section 2 we briefly describe the {\em Enzo-Galcon} code and our  
modeling of the ejection of magnetized IS plasma. The main results 
from the first {\em Enzo-Galcon} simulation are presented in section 3. 
We end with a brief discussion in Section 4.

\section{Simulation and Modeling}

To describe the evolution of ejected galactic magnetized gas we performed 
a high resolution simulation with the {\em Enzo-Galcon} code in the context 
of the $\Lambda$CDM cosmological model, with matter and dark energy density 
parameters $\Omega_m=0.27$, $\Omega_\Lambda=0.73$, respectively, and mass 
variance normalization $\sigma_8=0.9$ (taking $H_0 = 71$ $km\; s^{-1}\; 
Mpc^{-1}$). The code root grid includes $128^3$ cells which cover a comoving 
volume of $54^3 \, Mpc^3$ with two nested inner grids. The highest refined 
sub-grid covers a comoving volume of $27^3 \, Mpc^3$ divided into $128^3$ 
cells; this can be further refined adaptively by up to 5 levels, with a 
maximum $\sim$9 kpc resolution. We enforce maximum refinement in the 
vicinity of {\it galcons}, and 9 kpc is the resolution for gas exiting the 
galaxies. 

The simulation was initialized at $z=60$ and evolved until it was stopped 
at $z_r=3$, the ('replacement') redshift when early galaxies were already 
well developed and (as indicated by observations) star formation rate (SFR) 
peaked. At this time a halo-finding algorithm was employed to locate 89 
galactic halos with masses in the range $10^9-10^{12}\; M_\odot$ within a 
volume which eventually collapsed to form a rich cluster with total 
$M = 5.4 \times 10^{14} \; M_\odot$, and virial radius $R_V=1.7$ Mpc (ARN2). 
The baryonic contents of these halos were analyzed and replaced by {\it 
galcons}, with the baryon density profile in each {\it galcon} fit by a 
$\beta$ model. Both stellar and gaseous components are included in {\it 
galcons}, and since stars form in the same IS high gas density regions that 
contain most of the gas and can be assumed to have initially roughly similar 
spatial distributions, it is reasonable to approximate both by the same 
$\beta$-profile parameters, but with different central densities (see ARN2 
for details).

{\it Galcons} with these analytic density profiles are inserted into the 
centers of each halo, and assigned the halo velocity. Each {\it galcon's} 
central density and outer radius are determined from the fit and the value 
of the baryonic mass within the halo virial radius. An equal amount of 
baryons is removed from the simulated density field (without affecting the 
total mass density field, and thus preventing an unphysical instantaneous 
change in the simulated density distribution).

The mean initial baryonic mass density in galaxies can be determined by 
multiplying the mass density of halos from the Press \& Schechter (PS, 1974) 
mass function by the universal baryonic density parameter $\Omega_b$. The 
stellar mass density is calculated by integrating the cosmic SFR density (to 
be specified below) over the interval $[z_{i}, z_{r}]$. 

Upon initialization of the {\it galcons} the simulation was resumed and 
evolve to $z=0$. The motion of {\it galcons} was tracked using Enzo's N-body 
machinery. We followed mass and energy ejection processes - galactic winds 
and ram pressure stripping - through simple analytic models. Galactic winds 
reduce the total stellar mass while ram pressure stripping continuously 
reduces the {\it galcon} outer gas radius (as quantified below). Since 
galactic winds are SN driven, their elemental abundances are higher than in 
IS gas by a factor of $\sim 3$. Metal enrichment by each of the two processes 
is followed separately.

During cluster collapse and ensuing episodes of galaxy and subcluster 
mergers, IS media are partly stripped by tidal interaction between 
galaxies. As IC gas density builds up, ram-pressure stripping becomes 
increasingly more effective, especially in the central, higher density 
region. We quantify this hydrodynamical process by determining for each 
{\it galcon} (at all time steps) the stripping radius, where local IC gas 
pressure is equal to the local galactic IS pressure. It is simply 
assumed that all IS gas outside this radius is stripped in a relatively 
short dynamical time. [We generalize the analytic Gunn \& Gott (1972) 
stripping condition by including the contribution of DM to the galactic 
gravitational potential.] Observational evidence supports the 
expectation that stripping truncates the gaseous disk but does not 
modify the gas profile, nor appreciably affects the dynamics of the 
stellar and DM components of the galaxy (Kenney \& Koopmann 1999, 
Kenney, Van Gorkom \& Vollmer 2004). Thus, the outer radius of the 
{\it galcon} gas component is reduced to the stripping radius without 
modifying the central density or the scale radius of its profile. 

Galactic magnetic fields have a wide spectrum of scales over which 
the field changes direction. These `coherence' scales are in the range 
$l \sim 0.1\, - \, 10$ kpc, with the smallest `cells' reflecting local 
plasma conditions and processes, and the largest scale arising form the 
action of galactic dynamo (Carilli \& Taylor 2002) . In `normal' spiral 
galaxies values of the mean field on both these small and large scales 
are in the range $1-10\,\mu$G (Beck 2005); in our estimates we scale 
galactic fields to a characteristic mean (volume-averaged) value of 
$3\,\mu$G. Even though most cluster galaxies evolve from spirals to 
ellipticals (as an integral part of cluster formation and evolution), 
and partial stripping of their IS gas itself reflects this change, we 
note that mean field values in this range are typically deduced also 
in ellipticals (Mathews \& Brighenti 1997). When magnetized IS gas with 
such a spectrum of coherence scales is stripped the very high electrical 
conductivity in the plasma essentially ensures that fields are anchored 
to the outflowing gas (Rephaeli 1988). During the motion of a magnetized cell 
out of the higher density IS environment it expands and its density $n$ 
decreases; consequently, `flux freezing' implies that the mean field 
weakens according to $B \propto n^{2/3}$. 

Given that galactic fields are distributed in what are essentially 
separate regions with a spectrum of sizes, and that fields are frozen 
in the stripped gas, we simplify the treatment of the field transfer 
to IC space by assuming that cells are stripped out of their parent 
{\it galcon} essentially in tact. This is realistic for cells that are 
appreciably below a typical galactic size, $l \leq$10 kpc, namely for most 
(if not all) the range of coherence scales. While we do attain the highest 
level of spatial resolution, $\sim 9$kpc (implementing a geometric refinement 
capability of the {\it Enzo-Galcon} code) in the description of all the 
baryonic processes in {\it galcons}, this resolution limit is insufficient 
for following the evolution of cells with lower coherence scales. 
Consequently, we cannot address any aspect of the magnetic field dispersal 
process that is affected by the coherence scale. In fact, this limitation is 
the main reason for restricting our modeling of the field to mean scalar 
properties. 

The build-up of magnetic field energy density in IC space is described as 
follows: At each time interval the increase in the IC magnetic field energy 
density is calculated by subtracting its volume-weighted value at the 
beginning of the time interval from its value at the end of the interval. 
The product of the {\it galcon} initial velocity and the time interval 
provides a measure of the radial extent of a spherical shell across whose 
volume the field is re-distributed. This procedure is repeated until gas 
stripping out of {\it galcons} is negligible.

\section{Results}

\subsection{Evolution and distribution of IC fields}

The rise in the central and volume-averaged values of the field as 
magnetized plasma continues to accumulate in IC space is described in 
Figure 1. Late time evolution of the field is more rapid when 
ram-pressure stripping becomes increasingly more effective. 
Evolution of the field is essentially passive - no amplification 
or reconnection is assumed, simply because it is difficult to assess 
the effectiveness of these processes in clusters.

\begin{figure}
\centering
\epsfig{file=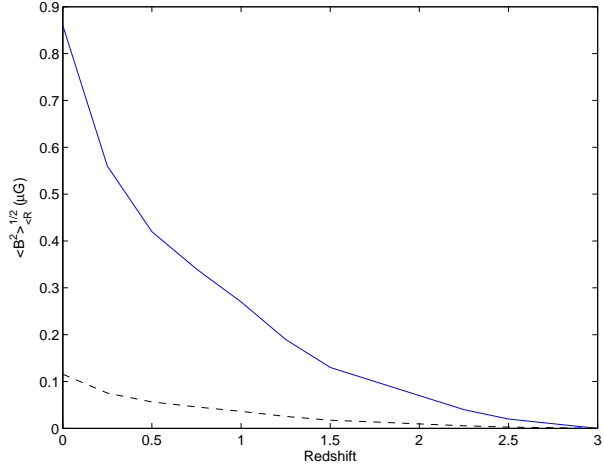, width=8cm, clip=}
\caption{Redshift evolution of the field central value (blue line), and its 
mean volume-weighted value over a region with radius of 1 Mpc (black dashed 
line).}
\end{figure}

The profile of the mean field, $<B^2>^{1/2}$, averaged over spherical shells, 
is
shown in Figure 2, with a central value $B_0 \simeq 0.9\,\mu$G 
(scaled to a mean galactic source field of $3 \,\mu$G). As we have previously 
shown (ARN2), the gas density and metallicity distributions can be well 
described by a profile with core radius in the $200-300$ kpc range, as 
observed in similar high-mass clusters (e.g., Sun et al. 2009). Similarly, 
we fit the profile of $<B^2>^{1/2}$ by a $\beta$ model, $B_{0}(1+
r^2/r_{B}^{2})^{-3\beta/2}$; the fit (shown in Figure 2) yields $r_B 
\simeq 350$ kpc, and $\beta \simeq 0.94$. 

\begin{figure}
\centering
\epsfig{file=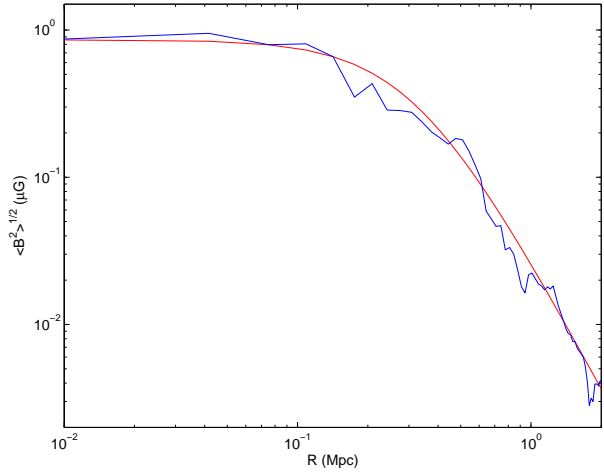,width=8cm, clip=}
\caption{Mean magnetic field as a function of radial distance from the 
cluster center (blue line) and the best fit to a 
$\beta$-profile (red line). The volume-weighted mean field is an 
emission-weighted average that was calculated assuming (electron) 
synchrotron emnission scales as $B^2$; see the text for more details.}
\end{figure}

\subsection{Comparison with observational results}

Lack of spatial information on IC fields limits comparison with observational 
results to just the mean field strength. The latter has been estimated from 
observations of radio synchrotron emission and FR measurements, for 
which we need to compute the respective field measures, volume 
average for the former, and a density weighted line-of-sight average 
for the latter. We emphasize that even with these measures comparison 
with observational results is not very meaningful due the fact that 
in observational analyses the field (and, in the case of sysnchrotron 
emission, also the electron density) are assumed not to vary across 
the cluster.

Measurements of IC synchrotron emission by relativistic electrons 
can be analyzed to determine the volume-average value of the mean 
(orientation-averaged) magnetic field, if a relation - such as 
equipartition - is assumed between the energy densities of particles 
(electrons and protons) and fields. If $p$ is the power-law index of 
the electron differential number density as function of the Lorentz 
factor, $n(\gamma) \propto \gamma^{-p}$, then the emitted flux scales 
as $B^{(p+1)/2}$, and since typical values of power-law indices of 
extended IC regions of radio emission are in the range $1-2$ (Rephaeli 
\ea 2008), $p \geq 3$. Therefore, the emission scales at least as 
steeply as $B^2$. This scaling is used in the calculation of the 
volume-weighted average of the field, shown in Figure 3. With our 
adopted $\beta$ profile for the gas density (and the assumption of 
flux-freezing), the volume-weighted mean field is nearly constant in 
the inner $\sim 100$ kpc region, beyond which it decreases steeply from 
a central value of $B_0 \simeq 0.9\,\mu$G to $\simeq 0.3\,\mu$G and 
$\simeq 0.1\, \mu$G at 0.5 Mpc and 1 Mpc, respectively. These estimates 
are in the typically deduced range of $0.1 - 1\, \mu$G (e.g., Ferrari 
et al. 2008), and are in agreement with estimates of the fields from 
joint analyses of radio and high-energy X-ray emission in several 
clusters (Rephaeli \ea 2008), if this emission originates in Compton 
scattering of the radio-emitting electrons by the CMB (Rephaeli 1979).

\begin{figure}
\centering
\epsfig{file=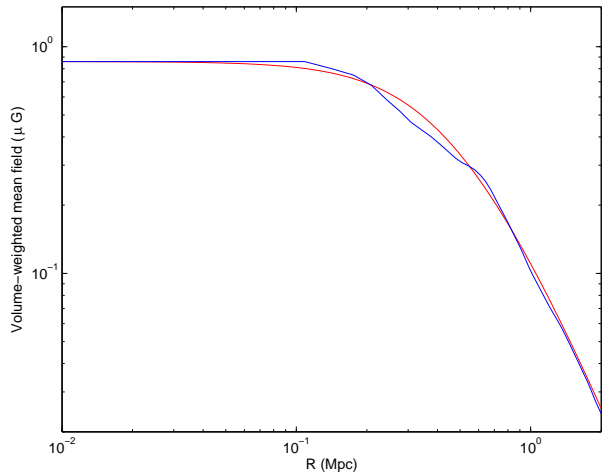,width=8cm, clip=}
\caption{Volume-weighted mean magnetic field as a function of radial 
distance from the cluster center (blue line), and its fit to a 
$\beta$-profile (red line).}
\end{figure}

The strength of IC fields has also been deduced from Faraday rotation of 
the polarization plane of radiation from radio galaxies located within, 
or lying behind clusters. Assuming a magnetic field morphology with cells of 
constant size, density, magnetic field strength, but with a random orientation 
inside each cell, the contribution to the rotation measure (RM) from each 
cell is given by 
\begin{equation}
RM = 812 \int^{0}_{L} n_e B_{||} dl \;\; (rad \; m^{-2}) \; ,
\end{equation}
where $n_e$ is the thermal electron density in $cm^{-3}$, $B_{||}$ is the 
line-of-sight magnetic field in microgauss, and $L$ is the path length in kpc. 
The precision with which the mean strength of IC fields can be determined is 
limited by their unknown morphology, by selective sampling of fields in the 
cluster core, and by considerable modeling and systematic uncertainties. There 
typically are only a few background radio galaxies located at different radial 
distances with respect to the cluster center, usually necessitating a 
statistical `stacking' analysis of a relatively small cluster sample. The 
spatial distribution of the field can only be inferred statistically from 
RM measurements of a large sample of clusters.

The distribution of RMs is generally patchy, indicating that large-scale 
magnetic fields are not regularly ordered on cluster scales, but have 
structures on scales as low as a few kpc. However, because the exact range 
of coherence scales is unknown, the interpretation of RM data assuming a 
constant scale over the entire cluster volume is clearly an oversimplification 
that introduces a substantial modeling uncertainty. Indeed, Newman et al. 
(2002) demonstrated that the assumption of a single-scale magnetic field 
leads to an overestimation of the magnetic field from RM data. 

To demonstrate this uncertainty, we calculated the RM for our simulated 
cluster using three different coherence scales spanning the range discussed 
in the literature: $\lambda=$9, 18, and 45 kpc. For each of these scales we 
computed an average magnetic field over all the cells that are contained 
in a region with a corresponding size, and selected a random orientation of 
the field, to account for the fields stochastic nature. The results are 
plotted in Figure 4, together with results from the statistical sample of 
RMs in 16 relaxed clusters from Clarke et al. (2001), and our best-fit to 
the measured values. While the comparison of our results with the composite 
profile (from 16 different clusters) has only limited meaning, the RM 
declining profile is roughly similar to the pattern seen in the data (even 
though some of the low-significance data at small radii are significantly 
higher than predicted by the model). Specifically, the best-fit line to the 
data matches well the predicted profile for a coherence scale of 
$\lambda \sim 18$ kpc. 

It is instructive to contrast our predicted mean field strength with that 
of Clarke et al. (2001), who derived a mean value of $\sim 5 \; \mu G$, 
assuming a coherence scale of 10 kpc. For our fiducial value $B_0 \simeq 
0.9\,\mu$G, the mean coherence scale that we deduce is roughly twice the 
value assumed by Clarke et al. This would seem to suggest that these 
results are roughly in agreement, since for a randomly oriented fields 
the mean field value deduced from FR measurements scales as $\lambda^{-2}$. 
Given the inherent indeterminacy in deriving both the mean field and the 
coherence scale from a single observable, the choice made by Clarke et al. 
is obviously not unique, even if admissible. While our predicted value for 
the mean field is, strictly speaking, a fiducial value based on scaling of 
the galactic source field to a value of $\sim 3\; \mu$G, this value is 
realistic and unlikely to be an underestimate by a more than $\sim 50$\%. 
Thus, our model provides a physical basis for realistically estimating the 
mean value of IC fields, in contrast to the much higher value adopted by 
Clarke et al. (2001), a value which is essentially based on an arbitrary 
choice for $\lambda$. 

In a more realistic description a range of values for the coherence 
scale should be taken for determining the RM distribution across the 
cluster. To do so we assume that the distribution of coherence scales 
can be well approximated by a Kolmogorov spectrum of cells with sizes 
in the range $[\lambda_1  - \lambda_2]$. Support for this assumption 
can be found in the recent simulations of Xu et al. (2010). In the 
context of our model of galactic origin for the fields this range is 
expected to correspond to the adiabatically expanded typical range in 
galaxies. Since the latter range is roughly a fraction of a kpc to few 
kpc, it is expected that the corresponding IC range is higher by a 
factor of $3-6$, based on $n^{-1/3}$ scaling of IC to IS plasma density 
ratio. Because the resolution limit in our simulated cluster is $9$ kpc, 
we sample the expected range by taking $\lambda_1=9$ kpc and $\lambda_2=45$ 
kpc. This is only a rough estimate intended solely for gauging the impact 
of taking a range of coherence scales (rather than just a single value). 

The predicted RM profile, shown in figure 4 (solid green line) is - not 
surprisingly - systematically higher than for the case of a single coherence 
scale with $\lambda < 45$ kpc. More importantly, the fact that this profile 
is above the best fit line to the RM data at (essentially) all radii, 
clearly demonstrates that the mean strength of the field need not be high 
for consistency with the RM measurements. This conclusion is only 
strengthened when we explicitly account for the considerable uncertainty 
in the data, and is also quite insensitive to the exact shape of the assumed 
spectrum, as long as it is a typical turbulent spectrum with a characteristic 
rise with scale ($\lambda$).

\begin{figure}
\centering
\epsfig{file=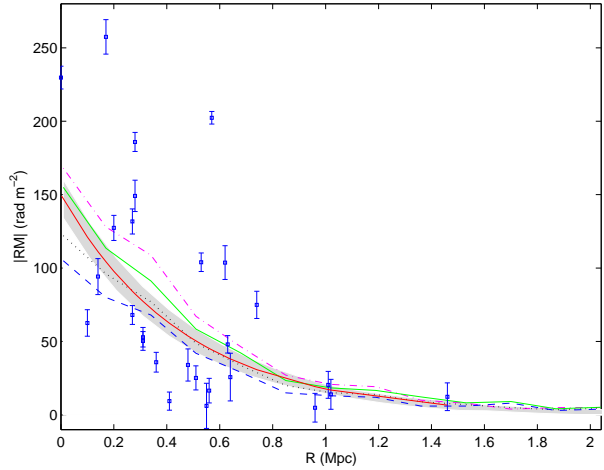,width=8cm, clip=}
\caption{Comparison of RM from the simulated cluster with observational 
data from 16 relaxed clusters from Clarke et al. (2001). We show the predicted 
RM profiles for coherence lengths of 9, 18 and 45 kpc by the blue dashed 
line, black dotted line and magenta dashed-dotted line, respectively. The 
green line is the RM curve calculated from a Kolmogorov coherence scale 
spectrum. The best fit to observational data and the 1-$\sigma$ uncertainty 
are shown by the red line and grey region, respectively.}
\end{figure}

\section{Discussion}

We present here results from cluster simulations with the {\em Enzo-Galcon} 
code incorporating a simple description of the ejection (by galactic winds) 
and ram-pressure stripping of magnetized IS media from cluster galaxies, 
and the direct implications on IC magnetic field. We show that magnetic 
fields anchored to the gas ejected and stripped out of galaxies can 
efficiently spread throughout the cluster volume, building a realistic 
declining spatial distribution for IC magnetic fields. In this approach we 
use a simple model for the origin of IC fields that does not invoke field 
amplification by mechanisms whose viability in the cluster environment is 
uncertain. Although field amplification may occur in some clusters it is 
questionable whether this can be generally the case in most clusters. We 
believe that the (relatively) minimalistic approach of a galactic origin 
for IC fields (Rephaeli 1988) provides a reasonable basis for gauging the 
main properties cluster fields.

Other contributions to IC fields could possibly be field amplification in 
small scale flux tubes (Ruzmaikin, Sokoloff \& Shukurov 1989) or during 
merger shocks (Roettiger, Stone \& Burns 1999). Magnetized plasma from 
radio galaxies (Kronberg et al. 2001) and AGN (e.g., Xu et al. 2009, 2010) 
could possibly contribute appreciably to IC the fields, but perhaps not 
ubiquitously, given the rarity of these sources as compared with normal 
galaxies. Xu. et al. (2009, 2010) carried out an MHD simulation with the 
{\it Enzo} code in order to follow the evolution of the magnetic field 
carried out by jets of a powerful AGN in the cluster center. Their 
simulation demonstrated that the ejected fields can be well spread 
throughout the cluster, and are amplified by gas turbulence during 
cluster evolution. The final field profile was found to be relatively flat, 
if AGN injection occurs before major mergers. AGN injection after major 
mergers results in more centrally peaked magnetic field distributions. 
But based on other MHD simulations (Dolag \ea 2001) it was concluded that 
in hierarchical formation of clusters, a correlation between RM and the 
X-ray surface brightness is expected to reflect the correlation between 
the magnetic field and gas density. Analysis of RM and X-ray brightness 
data for A119 led these authors to deduce that $B \propto \rho^{0.9}$, 
somewhat steeper than the slope we deduced in our model.

A more elaborate treatment of the evolution of IC fields was attempted by 
Donnert et al. (2008), who carried out a cosmological MHD simulation of 
clusters and explored a wide range of relevant parameters in galactic outflow 
models. While their approach is very different from our more minimalistic 
treatment, they too found out that IC fields generally have declining 
spatial profiles with predicted RM distributions that are in agreement with 
the observational results of Clarke et al. (2001).

The attractiveness of the galactic origin model for IC fields is further 
enhanced by the fact that it naturally yields relatively high $\sim 1 \, \mu$G 
field in the central region and the (very reasonable) expectation of an 
appreciably declining radial profile outside the core. These properties 
are in full agreement with measurements of radio emission and RM data. 
Moreover, the magnetic field energy within even $0.5$ Mpc (spherical) 
region is more than a hundred times smaller than for the case of a 
constant $\sim 5 \,\mu$G field.

Given the rudimentary level of our understanding of magnetic fields and 
nonthermal phenomena in clusters, more comprehensive observational and 
theoretical studies are needed in order to determine the full range of 
possible sources of IC magnetic fields and nonthermal particles. Extensive 
radio measurements of many more clusters are needed to obtain more detailed 
information on the spatial distributions of the emission and RMs. Our 
simulation work will continue with the considerably improved 
{\em Enzo-Galcon} code.

\section*{ACKNOWLEDGMENTS}

\noindent Work at Tel Aviv University is supported by US-IL Binational Science 
foundation grant 2008452. The simulations were performed on the Data 
Star system at the San Diego Supercomputer Center using LRAC allocation 
TG-MCA98N020.

\end{document}